\documentclass[%
 reprint,
 amsmath,amssymb,
 aps,
nofootinbib,
]{revtex4-2}

\usepackage{graphicx}
\usepackage{amsmath}
\usepackage{amssymb}

\usepackage{xcolor}

\begin{document}


\title{Nonaffine motion in flowing highly polydisperse granular media}

\author{Pablo Eduardo Illing}
\email{pablo.eduardo.illing@emory.edu}

\author{Eric R. Weeks}%
\email{erweeks@emory.edu}
\affiliation{Department of Physics, Emory University, Atlanta, GA 30322, USA
}%


\date{\today}

\begin{abstract}
We study the particle-scale motion of highly polydisperse hard disks flowing in a two-dimensional bent channel. We use various size distributions of particles, in which the largest particles are up to five times larger than the smallest.  The disks are pushed through an L-shaped channel to drive the particle rearrangements.  Although the mean flow is essentially independent of the polydispersity, the motion of individual particles becomes more nonaffine on average for higher polydispersity samples.   We characterize the nonaffine motion, finding a qualitative difference in the behavior of small and larger particles: the smaller disks have more nonaffine motion, induced by the larger particles.
\end{abstract}

\maketitle

\section{\label{sec:level1}Introduction}

The flow of soft amorphous materials has been a subject of extensive study for decades \cite{falk_dynamics_1998,nicolas_Deformation_2018,durian_Foam_1995,tong_Crystals_2015,hebraud_yielding_1997,petekidis_rearrangements_2002,schall_structural_2007}. These amorphous materials are common, both in nature and in industry, in the form of colloids, emulsions, granular media, foams, and food products, among others \cite{chen_microscopic_2010,besseling07,friberg_Foams_2010,gonzalez-rodriguez_Soft_2012,schadle_characterization_2022}. Much prior research on the flow and rheological properties of these materials examined systems with particles or droplets of similar sizes and properties \cite{mason_Yielding_1996,hebraud_yielding_1997,schall_structural_2007,chen_microscopic_2010,besseling07,vasisht_rate_2018,tsai_signature_2021,yamamoto_nonlinear_1997,olsson_critical_2007,utter_experimental_2008,lemaitre_rate-dependent_2009,manning_vibrational_2011,cubuk_identifying_2015,losert_particle_2000,patinet_connecting_2016}.  In contrast with these model systems, many natural systems and materials have components with a range of sizes. For example, the size ratio between the largest and smallest particles can be a factor of 10 or more \cite{colwell_lunar_2007,guber_effect_2003,caracciolo_raindrop_2012,burton_quantifying_2018,haeberli_permafrost_2006}. The presence of disparate-sized particles is known to affect the flow behavior of various systems, including sand and gravel deposition \cite{gladstone_experiments_1998,ahfir_porous_2017,dorrell_polydisperse_2013,werner_eolian_1995}, hopper flow \cite{gundogdu_discharge_2004,govender_study_2018,vlachos_investigation_2011}, and geophysical phenomena such as avalanches, land slides, and glacier flow \cite{pitman_two-fluid_2005,viroulet_kinematics_2018,tsai_signature_2021,amundson_quasi-static_2018,amundson_ice_2010,burton_quantifying_2018}.  Mixtures of various size components can also determine consistency and texture in food products \cite{schadle_characterization_2022,taylor_shear_2009}.  Having a mixture of sizes in a sample is termed ``polydispersity.''

The study of these kinds of systems has led to interesting physical behavior when compared to their monodisperse counterparts. As an example, polydisperse hard spheres can phase separate into multiple crystalline phases \cite{sollich_crystalline_2010}.  In active matter, polydispersity leads to the emergence of new phases \cite{kumar_effect_2021}. Experimental and computational studies on the compression and stretching of particle rafts have shown that polydispersity greatly affects their structural properties, such as their compressional yielding threshold \cite{ono-dit-biot_Rearrangement_2020,illing_compression_2024}.
In granular materials, force chains become drastically more heterogeneous in more polydisperse systems, affecting the material's jamming point and rheological properties \cite{nguyen_effect_2014,nguyen_effects_2015,cantor_rheology_2018,papadopoulos_Network_2018}. In particulate suspensions, the polydispersity of the particles strongly impacts the viscosity of the suspension: for example, adding small particles can lower the viscosity \cite{pednekar_bidisperse_2018}.

Previous studies of sheared soft materials typically wish to avoid crystalline order, so often a bidisperse mixture of particles is used, or a single type of particle with mild polydispersity \cite{schall_structural_2007,utter_experimental_2008,yamamoto_nonlinear_1997,hebraud_yielding_1997,petekidis_rearrangements_2002,schall_structural_2007,chen_microscopic_2010,vasisht_rate_2018,tsai_signature_2021,cubuk_identifying_2015,olsson_critical_2007,lemaitre_rate-dependent_2009,manning_vibrational_2011}. 
Polydispersity $\delta$ is defined as the standard deviation of the particle radii divided by the mean radius.  A frequently studied system is bidisperse, with equal numbers of small and large particles with size ratio $1:1.4$, yielding $\delta = 0.17$ \cite{perera99,speedy99,ohern_Jamming_2003,olsson_critical_2007,yamamoto_nonlinear_1997}.

Two prior studies examined highly polydisperse emulsions ($\delta \geq 0.5$), with the size ratio between the largest and smallest particles as large as $10:1$ \cite{jiang_effects_2023,clara-rahola_affine_2015}.  These studies found that large and small particles play different roles in the flow of the sample, with large particles moving more smoothly, while small particles move more erratically.  This has implications for how particles are mixed and also consequences for the rheological response: highly polydisperse systems have well-mixed small particles and are easier to flow \cite{jiang_effects_2023}.  These two studies only considered emulsion droplets at high volume fractions (above jamming); because the droplets are soft they can still flow, but leaving unanswered the question as to whether these prior observations generalize to hard particles at packing fractions below jamming.

In this paper, we show the effect the particle size distribution has on the flow of granular materials, in particular, how individual particle motions deviate from the mean flow pattern, resulting in local rearrangements.  The granular particles used for this work are hard acrylic disks. We use 11 different particle size distributions with varying polydispersity, ranging from $\delta = [0.2, 0.48]$.  We push mixtures of these disks through an L-shaped channel to cause the particles to rearrange, and study individual particle motions during this flow. We find that large particles are more likely to follow the mean flow, and more likely to perturb the motion of nearby smaller particles so that the latter do not follow the mean flow.  Our observations confirm the prior understanding \cite{clara-rahola_affine_2015,jiang_effects_2023}, extending those observations to hard particles.  We additionally find that the influence of the perturbation from the larger particles extends only a short distance from the surface of the large particles, about 2-3 small particle diameters.

\section{Experimental Methods}

\subsection{Particles and Flow Chamber}

\begin{figure}
\includegraphics[scale=0.16, origin=c]{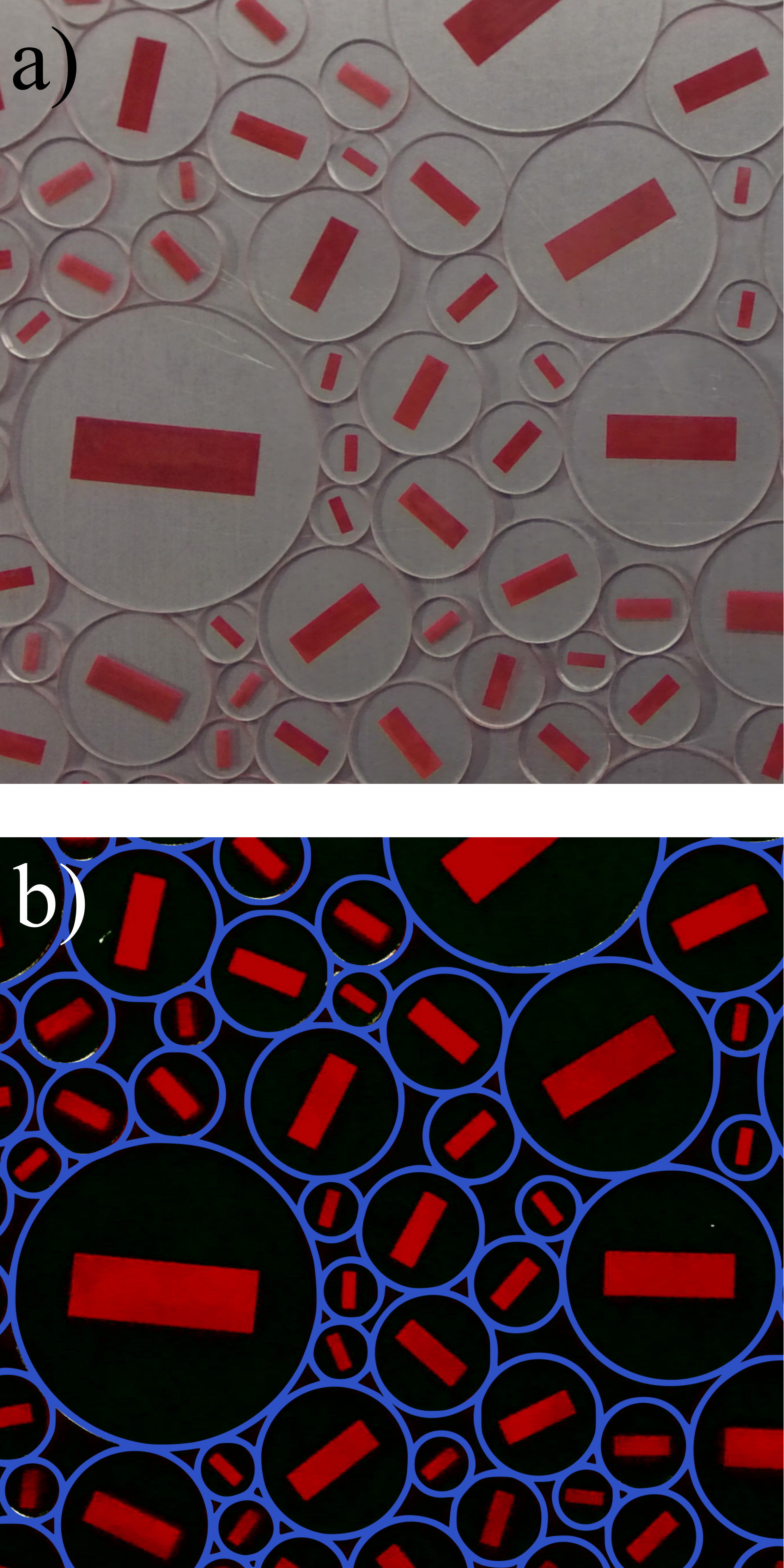}
\caption{\small Sample image of the C3 distribution illustrating the process used for image analysis and particle tracking. (a) Photograph of the experimental device with particles in place, under normal lightning conditions. (b) Photograph of the same frame using green/yellow lightning to highlight the red rectangles painted on the particles. We have superimposed blue rings to illustrate the results of our particle tracking. The largest particle  on the left has a diameter of $5.7$ cm, with its rectangle being $3.5$ cm wide. For more details on the particle distributions refer to Table~\ref{tab:sd} and Fig.~\ref{fig:partcount}.}
\label{fig:screenshot}
\end{figure}

Our samples are composed of circular disks, cut from 2.9~mm thick cast acrylic sheets with a laser cutter.  Frosted rectangles are etched into the center of the disks during the cutting procedure to facilitate tracking the particles' positions and orientations. In practice, we do not see any interesting results regarding particle rotation, so we ignore the orientational information for this paper.  As a final step, the frosted rectangles are painted red using a felt-tip marker, to make particle tracking easier. A sample of the particles is shown in Fig.~\ref{fig:screenshot}(a). Here we can see a wide variety of particles with radii ranging from $0.635 ~\mathrm{cm}$ to $2.85 ~\mathrm{cm}$. The thickness of the particles and the smallest radius have been chosen so that the particles do not tip over when pushed.

\begin{figure}
\includegraphics[scale=1.4, origin=c]{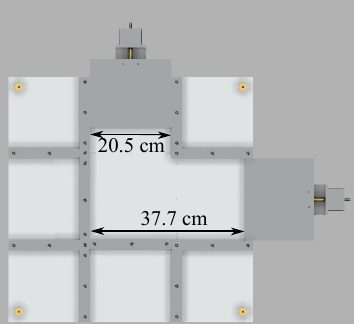}
\caption{\small Top down schematic of the experimental device, set up in the "L" configuration. Plungers have been placed at each end of the flow geometry, this allows us to push the particles back and forth. The plungers both move at a set speed $v_p=0.22 \mathrm{cm/s}$ and maximal extension of $10.7~\mathrm{cm}$.}
\label{fig:devic_pic}
\end{figure}

The experimental device we use for this project consists of a large square aluminum base, with each side measuring $53.3~\mathrm{cm}$, on which we can screw in several divisions. This work will focus on the `L' configuration with the dimensions given in Fig.~\ref{fig:devic_pic}; this geometry is similar to some prior work \cite{desmond15}. This configuration consists of a track of total length $75.4 ~\mathrm{cm}$, bent at a right angle at the $37.7~\mathrm{cm}$ mark, and with a $20.5~\mathrm{cm}$ width opening.  To push the particles through our experimental setup at a steady velocity, we use mechanical plungers, with a set speed of $v_p=0.22~\mathrm{cm/s}$ and a maximal extension of $10.7~\mathrm{cm}$.  Particles are added to fill the available area in the flow channel, with the number of particles ranging from 164 to 370 depending on the size distribution used.

In order to ensure a two-dimensional flow of the disks confined to the surface of the device and prevent the particles from flowing over each other, a transparent acrylic sheet is screwed on top of the experimental apparatus. This sheet can be easily removed to facilitate the placement of particles in the device.

To record the particles, we use a MOKOSE UC70 color camera with $2100 \times 2100$ pixel resolution, operating at 6 frames per second.  The camera is placed directly above the experimental device.  We light the experiment using an array of colored LED lights, diffused through a screen, to achieve a homogeneous light source. The LED lights are set to a yellow/green color, to make the red rectangles on the particles contrast better against the reflective aluminum background.

We also measure the relevant friction coefficients in the experiment.  The friction coefficients between the particles and the aluminum base are measured by tilting the surface until the disks move, and measuring the subsequent velocity of the sliding disks.  We obtain $\mu_{\mathrm{static}} \approx 0.5$ and $\mu_{\mathrm{dynamic}}\approx 0.3$. The value of $\mu_{\mathrm{dynamic}}$ combined with the low plunger speed we use allows us to calculate the stopping time at $\Delta t_\mathrm{stop}\sim 7\times 10^{-4} ~\mathrm{s}$.  Essentially, when the plunger stops moving, particles stop instantaneously; inertia is negligible.  The friction coefficients between the acrylic particles themselves are also measured: $\mu_{\mathrm{static}}\approx 0.4$ and $\mu_{\mathrm{dynamic}} \approx 0.3$.

\subsection{Experimental procedure}\label{SS:Exp_proc}

\begin{figure}
\includegraphics[scale=0.3, origin=c]{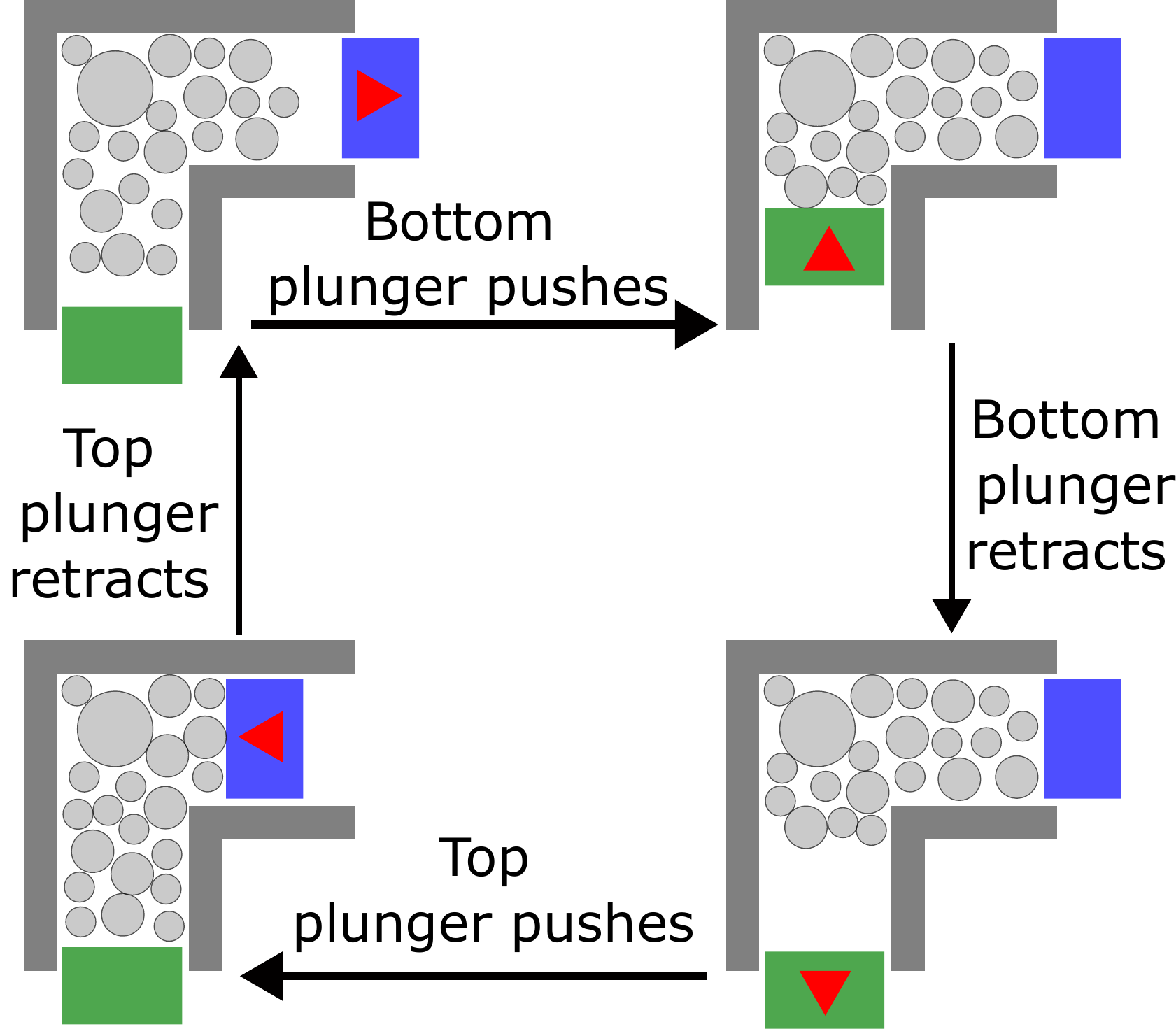}
\caption{\small A top down sketch of the experimental device, setup in the ``L" configuration. Plungers have been placed in such a manner as to block the outlets and study the sloshing back and forth of a  given particle configuration. 
 The triangles indicate which plunger has moved at a given stage.}
\label{fig:exp_procedure}
\end{figure}

To initialize the experiment, the particles from a chosen size distribution are randomly placed in the flow channel, which we then cover with the previously mentioned acrylic sheet. 
The particles are then pushed through the experimental device in a cyclical manner; a diagram explaining the experimental process is shown in Fig.~\ref{fig:exp_procedure}.  We start with one fully retracted plunger to make space for particles to move, and then the other plunger pushes the particles through the channel until they almost reach the retracted plunger.  We then retract the previously moving plunger to make room for the next cycle, and then the particles are pushed in the opposite direction by the previously stationary plunger.  In this way, the particles are forced back and forth around the L-shaped channel.  We repeat this 10 times per experimental run.  During the first two cycles, we observe that the area fraction rises nontrivially due to particle rearrangements.  Accordingly, we only analyze the final eight cycles for which the area fraction has reached a steady state.  Once a run is completed, we remove the top cover, randomize the position of the particles using a random number generator for the positions, and carry out the next run of the experiment using the same procedure.  For each size distribution, we carry out five runs, each with randomized initial positions.

\subsection{Particle size distributions}

Our aim is to study the effect of the particle size distribution $P(R)$ on the flow of our particles.  The quantity which characterizes the size variety is the polydispersity $\delta$:
\begin{equation}
    \delta= \sqrt{\langle \Delta R^2 \rangle/}\langle R \rangle
    \label{eq:polydipersity}
\end{equation}

\noindent where $R$ is the radius of a given particle, $\Delta R=  R-\langle R \rangle$, and the moments of $R$ (and $\Delta R$) are given by $\langle R^n \rangle=\int R^n P(R) dR$ (and $\langle \Delta R^n \rangle=\int \Delta R^n P(R) dR$).

\begin{table}[!bthp]
\centering
\begin{tabular}{|c|c|c|c|}
    \hline
    {\small{Size Distribution}} & {Size Ratio} & {Number Ratio} & $\delta$\\
    \hline
     {Bidisperse} &   {\small{3:2}} & {\small{1:1}} &   $0.20$ \\  
  \hline
      {T1}  &   {\small{4:3:2}}&  {\small{1:8:8}} & $0.23$ \\  
    \hline
      {T2a}  &   {\small{6:3:2}}&  {\small{1:24:24}}& $0.27$  \\  
     \hline
      {T2b}  &   {\small{6:3:2}} &  {\small{1:18:18}} & $0.29$ \\ 
    \hline
      {T2c}  &   {\small{6:3:2}}&  {\small{1:11:11}}& $0.32$  \\  
    \hline
      {T2d}  &   {\small{6:3:2}}&  {\small{1:9:9}}& $0.35$ \\  
    \hline
      {T3}  &  {\small{8:3:2}} & {\small{1:26:26}} & $0.35$  \\ 
    \hline
      {T4}  &   {\small{10:3:2}}&  {\small{1:28:28}}& $0.42$  \\  
    \hline
      {C1}  &  {\small{6 to 2}} &  {\small{1 to 25}}& $0.31$  \\  
    \hline
      {C2}  &   {\small{8 to 2}}&  {\small{1 to 50}}& $0.40$  \\  
    \hline
      {C3}  &   {\small{10 to 2}}&  {\small{1 to 55}}& $0.48$ \\  

    \hline

\end{tabular}
\caption{The disk size distributions used. The first column shows the name of each distribution, with ``T'' standing for tridisperse and ``C'' standing for an approximation to a continuous size distribution.  The second and third columns give the size and number ratios of the disks used, respectively. The fourth column gives polydispersity $\delta$ of each distribution. For the C1 size ratios are 2:3:4:5:6; for C2 the size ratios are 2:3:4:5:6:7:8; and for C3 the ratios are 2:3:4:5:6:7:8:9:10. For more information about the number ratios of C1, C2, and C3, Fig.~\ref{fig:partcount} shows a histogram with the amount of particles of each size in these distributions.}
\label{tab:sd}
\end{table}

\begin{figure}
\includegraphics[scale=0.61, origin=c]{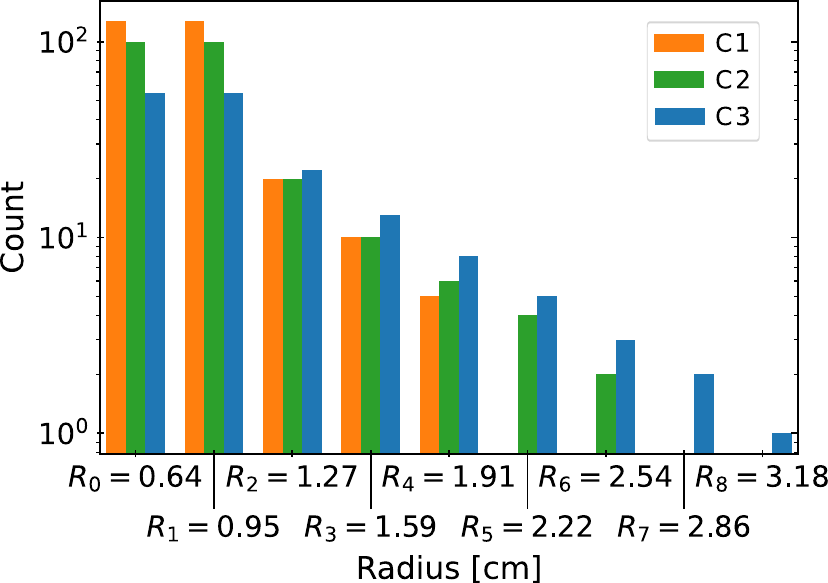}
\caption{\small Histogram for the three continuous distribution of particles, C1, C2 and C3. These distributions are built using particles sizes ``continuously," starting from $R_0$ and $R_1$.}
\label{fig:partcount}
\end{figure}

Specifically, there are a total of nine different radii that we use.  In terms of $a=0.3175$~cm (an eighth of an inch), the smallest particles have a radius  $R_0=2a=0.635$~cm, and the other radii are defined by $R_n=(n+2)a$, up to $R_8=3.175$~cm.  Table \ref{tab:sd} contains information on the size ratio, number ratio, and polydispersity of each size distribution, which are composed of subsets of these particles.  The simplest size distribution is the bidisperse distribution with equal numbers of $R_0$ and $R_1$ sized particles, which avoids hexagonal ordering \cite{meer24}.  This distribution is similar to the ``canonical" $1:1.4$ size ratio often studied in previous works \cite{perera99,speedy99,ohern_Jamming_2003,yamamoto_nonlinear_1997,olsson_critical_2007}.  
The tridisperse distributions, labeled T1 through T4, are built using the bidisperse distribution with an added third particle species of greater size. These are useful for probing the effect of changing the polydispersity (up to $\delta = 0.42)$, or for distributions T2d and T3 cases, fixing $\delta=0.35$ but changing the largest particle size.  The three ``continuous'' size distributions are built from a range of discrete particle sizes; Fig.~\ref{fig:partcount} shows a histogram of the particle counts for these distributions.  They vary by the largest included size, which also affects their polydispersity, as given in Table~\ref{tab:sd}.

\subsection{Image Analysis}\label{SS:mp}

The first step in our image analysis is to split each frame of a recording into its corresponding red/green/blue values. We subtract the green channel from the red channel, which results in the red rectangle of each particle being strongly highlighted against a dark background.  We threshold the resulting image and identify all groups of connected pixels above the threshold.  From each group of pixels, we find the center of mass, area, and aspect ratio.  Knowing that the valid features are rectangles of specific known sizes allows us to filter out falsely identified particles, as is frequently done in particle tracking \cite{crocker_Methods_1996}. After we have identified the particles in each frame, we track their trajectory using standard software \cite{crocker_Methods_1996}. In Fig.~\ref{fig:screenshot}(a) we show an example of a raw image of our particles using natural lighting.  Figure \ref{fig:screenshot}(b) shows the same particles but using the lighting conditions described above and post-processing the image to highlight the rectangles. In Fig.~\ref{fig:screenshot}(b), we also show superimposed rings from the results of our particle identification method.

\section{Results}
\subsection{Mean Flow}
\label{sub:bkgrnd}

\begin{figure}
\includegraphics[scale=1, origin=c]{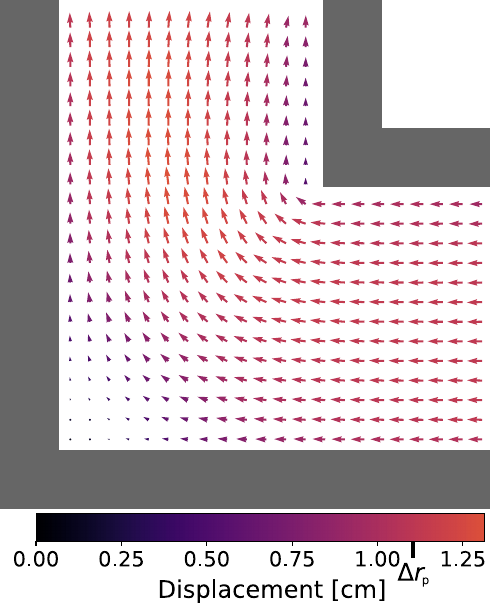}
\caption{\small Mean displacement field $\Delta \vec{r}_\mathrm{mean}(x,y)$.  The displacements are calculated using $\Delta t =5\mathrm{s}$. The bold line on the scale bar indicates $\Delta r_\mathrm{p}= v_p \Delta t =1.1 \mathrm{cm}$, the displacement corresponding to the plunger motion over $\Delta t$. On the lower right inlet, we observe a plug like flow closer to the plunger, which then turns into a more shear like flow as the particles turn the corner. Closer to the lower left corner we see an area with almost no displacements, signified by the shorter arrows with the darker color.}
\label{fig:bkgrnd_flow}
\end{figure}

We first consider the mean flow properties of our samples. We start by calculating the displacements of the particles using a time scale $\Delta t = 5$~s which corresponds to roughly a tenth of the total duration of a cycle (the disks moving in one direction as one plunger pushes on them).  During this time interval, the plunger moves $\Delta r_{\rm p} = v_{\rm p} \Delta t = 1.1$~cm, slightly less than the diameter of the smallest particles ($1.27$~cm).  We then spatially bin the data with a resolution $\Delta w = 1.6~\mathrm{cm}$, which is the mean diameter of the two particle species in the bidisperse sample.  Within each bin, we find the mean displacement vector, averaging over all particles and all times.  To compute this average, we also exploit the symmetry of the back-and-forth motion in the ``L'' (Fig.~\ref{fig:exp_procedure}), and thus reorient the data so that the active plunger is always at the lower right corner. The result is the vector field $\Delta \vec{r}_\mathrm{mean}(x,y)$ shown in Fig.~\ref{fig:bkgrnd_flow}. We see plug-like flow on the lower right inlet, corresponding to the active plunger, which enforces that all particles contacting it move with the plunger velocity $V_{\rm p}$.  In the corner region of the ``L'' particles change the direction of their motion.  The particles in the lower left corner barely move on average, whereas the particles in the middle of the corner region move significantly, resulting in a velocity gradient.  In the upper outlet region, the motion of the particles is slightly slower near the sidewalls.  To conserve particle flux, this means that particles near the center of the outlet region are moving slightly faster than the plunger speed.

Note that Fig.~\ref{fig:bkgrnd_flow} is averaged over all particle size distributions.  We separately compute the mean flow field for each particle size distribution and find that the different flow fields are nearly the same within the noise, with no systematic variation.  Accordingly, to reduce the noise, we consider $\Delta \vec{r}_\mathrm{mean}(x,y)$ as a useful reference mean flow for all experiments.

\subsection{Nonaffine displacement and local particle rearrangement}
\label{sub:dt15}

\begin{figure*}
\includegraphics[scale=0.93, origin=c]{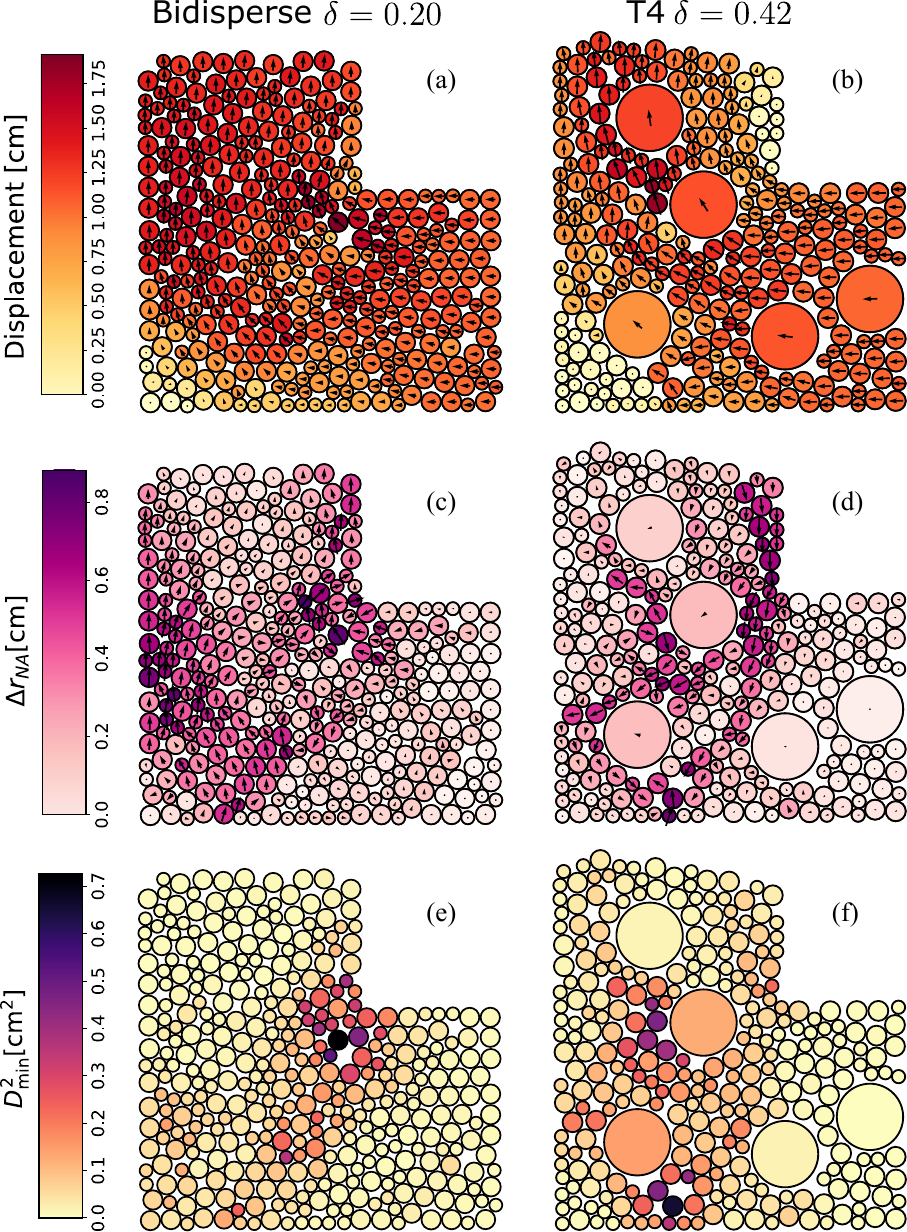}
\caption{\small {Snapshots showing the different quantities measured in the flow for the bidisperse size distribution (left column panels) and the T4 size distribution (right panels). From top to bottom, (a,b) display the displacement of each particle, (c,d) show the nonaffine displacement, and (e,f) show $D^2_{\mathrm{min}}$ for each disk.  The plunger at lower right enforces that contacting particles move with velocity $v_p$, and thus have displacements $\Delta r_p=1.1 \mathrm{cm}$.  For the bidisperse distribution, regions of higher nonaffine motion ($\Delta r_{\rm NA}$ or $D^2_{\rm min}$) are typically associated with locations of higher strain.  For the tridisperse distribution, the largest particles generally have less nonaffine motion, but nearby smaller particles often have more nonaffine motion.  For both distributions, the flow near the moving plunger (lower right) is plug flow: 
thus no shearing and no nonaffine motion.}}
\label{fig:3x4}
\end{figure*}

Of course, the flow field shown in Fig.~\ref{fig:bkgrnd_flow} is averaged over all particles and all times; at any specific moment, individual particles are often found moving in different directions, and it is only their average which is a smooth function of space. As an example, we can look at Figs.~\ref{fig:3x4}(a,b) which show the displacement of particles for two different distributions, bidisperse and T4, for a single frame.  The T4 distribution is constructed by adding five large particles to the bidisperse distribution, where the large particles are five times larger than the smallest particles.  In Figs.~\ref{fig:3x4}(a,b), the length of the arrow and also the color of the particle signify the magnitude of the displacement, with darker colors for larger displacements.

Starting with the displacement of bidisperse case shown in Fig.~\ref{fig:3x4}(a), we observe many of the characteristics highlighted in Fig.~\ref{fig:bkgrnd_flow}.  These characteristics include plug-like flow near the plunger inlet and shear-like flow on the outlet side.  Of course, Fig.~\ref{fig:bkgrnd_flow} shows the mean displacement field $\Delta r_\mathrm{mean}(x,y)$ which is an average over data such as Fig.~\ref{fig:3x4}(a).  At the specific time shown in the latter, the lower left corner has very small displacements, with some particles at the corner being completely still.  There are also a few particles with displacements larger than that of the plunger, $\Delta r_{\mathrm{p}}=1.1$~cm. Another interesting feature is the difference in displacements between neighboring particles: in contrast to Fig.~\ref{fig:bkgrnd_flow} where the colors change smoothly as a function of space, for the discrete particles there are instances in Fig.~\ref{fig:3x4}(a,b) where a region has a mixture of colors.

To quantify these behaviors of individual particles at individual moments in time, we consider nonaffine displacements of the particles, $\vec{\Delta r}_{\mathrm{NA}}$.  Affine motion occurs when the particle displacements are a smooth function of their positions.  To define the nonaffine motion of our particles, we subtract the displacement calculated from the time- and particle-averaged flow field from the displacements $\vec{\Delta r_i(t)}$ of specific particles $i$ at specific times $t$:
\begin{equation}
    \Delta \vec{r}_{i\mathrm{NA}}(t)=\Delta\vec{r}_i (t) - \Delta \vec{r}_{\mathrm{mean}}(x,y).
    \label{eq:non_aff}
\end{equation}

\noindent Here $\Delta \vec{r}_{\mathrm{mean}}(x,y)$ is the mean displacement at the initial position $(x,y)$ of particle $i$.  Similar measures of nonaffine motion have been used in previous work to characterize flow in amorphous soft materials
\cite{leonforte05,lemaitre07,maloney08,besseling07,chen_microscopic_2010,clara-rahola_affine_2015,jiang_effects_2023,jiang_isomorphs_2023}.

Figures~\ref{fig:3x4}(c,d) show a vector map for the nonaffine displacements for the data corresponding to panels (a,b). As we did for Figs.~\ref{fig:3x4}(a,b), the arrows are the nonaffine displacements, and darker particle colors indicate a larger magnitude of $\Delta r_{\mathrm{NA}}$. We see that most particles with high nonaffine displacements occur closer to the central area and close to the walls in the upper section. In the bidisperse flow this occurs due to the rearrangement of particles as the flow changes from plug-like flow to shear-like, and the particles need to navigate the turn around the corner. For the more polydisperse sample, this behavior is still present.  However, the flow is also disrupted by the larger particles.  This is seen as most instances of nonaffine motion now occur around the larger particles, other than the lower right region where there is plug-like flow. Another interesting note is that while large particles cause a disruption in the surrounding flow, these large particles themselves have small $\Delta r_{\mathrm{NA}}$ values compared to their neighbors.

A simple way to explain this behavior is to picture a large particle as it moves around the corner.  The mean flow is shown in Fig.~\ref{fig:bkgrnd_flow}, and for a sufficiently large particle near the top right corner, it would exist in regions where the mean flow changes both in magnitude and direction.  Given that the large particle feels forces from a variety of adjacent smaller particles, it makes sense that the large particle will, on average, still follow the mean flow expected for the large particle's center.  However, any nearby smaller particles will try to move according to the local flow field they experience. If a smaller particle is close to the larger particle but in a normally faster section of the flow field, the smaller particle will need to move around the larger particle. In contrast, a small particle in a slower local flow will be pushed out of the way by the larger particle \cite{jiang_effects_2023}.

Of course, it is possible that large particles could locally induce a smooth flow of themselves and their neighboring particles.  To look for this, we consider an alternate definition of nonaffine motion introduced by Falk and Langer in 1998 \cite{falk_dynamics_1998} and widely used since then \cite{hassani_probing_2019,utter_experimental_2008,chen_microscopic_2010,priezjev17,cubuk_identifying_2015}.

The key idea is to examine a local group of particles and fit their displacements to a strain tensor using a least-squares fit.  The least squares fit error, $D^2_{\rm min}$, then quantifies the extent to which that local group of particles is {\it not} well described by a simple strain tensor, and thus serves as a measure of the nonaffine motion of that group of particles.  To compute this quantity, we select a particle $n=0$ and a set of its nearest neighbors and fit the displacements of all of these particles at a specific time $t$ to a local strain tensor $\epsilon_{ij}$.  The fitting is least squares where we find $\epsilon_{ij}$ to minimize the quantity
\begin{equation}
   \begin{split}   
    D^2(t,\Delta t)=\sum_n \sum_i \bigl\{ r^i_n(t)-r^i_0(t)-\sum_j(\delta_{ij}+\epsilon_{ij})\\ \times[r_n^j(t-\Delta t) -r^j_0(t-\Delta t)]  \bigr\} ^2,
    \end{split}
    \label{eq:D2min}
\end{equation}
\noindent where $n$ indexes the neighbors of the reference particle with the index $n=0$. The indices $i,j$ refer to the spatial coordinate components, $\epsilon_{i,j}$ is the best least-squares fit strain matrix characterizing the region, and $\delta_{ij}$ is the Kronecker delta.  The residual error after least squares fitting, $D^2_{\rm min}$, is our measure of local nonaffine motion \cite{falk_dynamics_1998}.  Here, rather than defining the affine flow through the space- and time-averaged flow, the affine flow is determined locally in space and time.

For random packing of highly polydisperse particles $D^2_{\mathrm{min}}$ is strongly dependent on the number of nearest neighbors included in the sum over $n$ in Eq.~\ref{eq:D2min}: more neighbors allow for more deviations from the mean strain matrix, increasing $D^2_{\mathrm{min}}$ \cite{jiang_effects_2023}.  Following prior work by Jiang {\it et al.} \cite{jiang_effects_2023}, we use the $N_{\mathrm{nbs}}=15$ closest particles as the nearest neighbors, defining the distance between particles as surface-to-surface:
\begin{equation}
    d_{0n}=|\vec{r}_n-\vec{r}_0|-R_n-R_0
    \label{eq:distance}
\end{equation}
\noindent where, as before, $\vec{r}_0$ is the position of the reference particle, $\vec{r}_n$ the position of neighbor $n$, and $R_0$ and $R_n$ their respective radii. Using Eq.~\ref{eq:distance}, particles in contact are at distance $d_{0n}=0~\mathrm{cm}$.  This definition allows us to fairly compare the $D^2_{\mathrm{min}}$ of particles of different sizes and across multiple size distributions.  Using $N_{\mathrm{nbs}}=15$ guarantees a full layer of neighbors around the largest particles and roughly two layers of neighbors for the particles in the bidisperse case.  


\begin{figure}
\includegraphics[scale=0.59, origin=c]{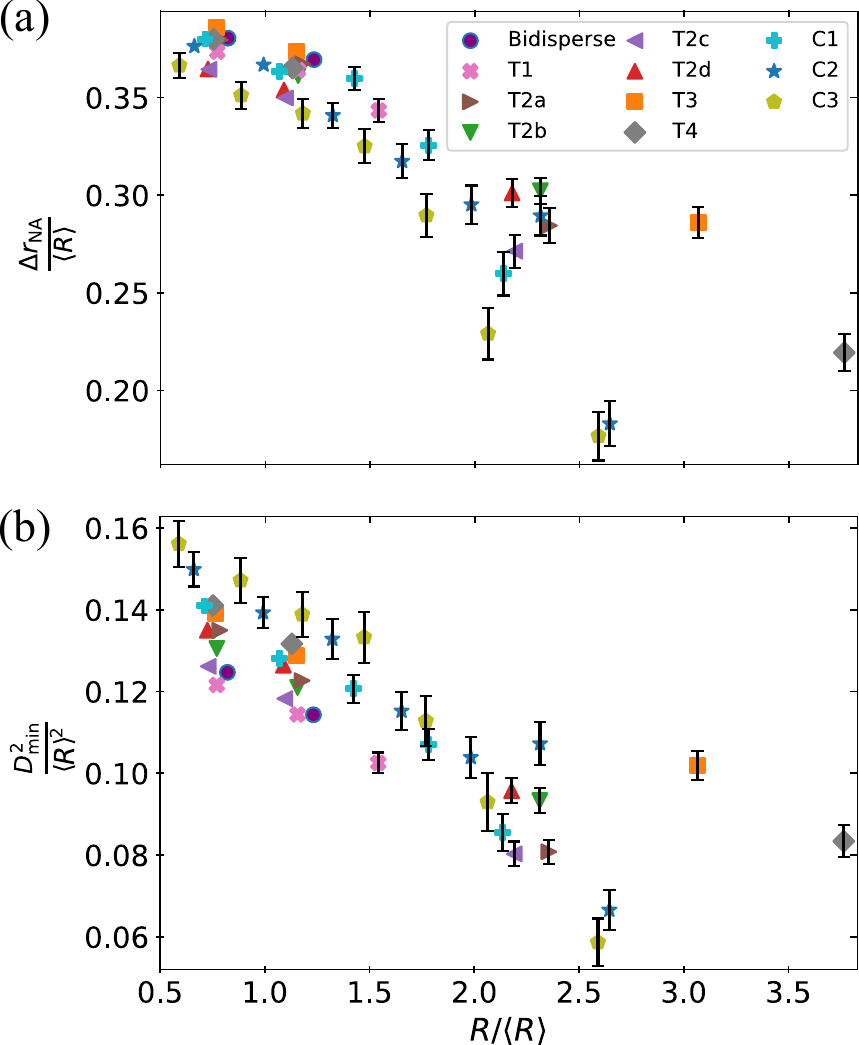}
\caption{\small (a) The mean magnitude of the nonaffine motion $|\Delta r_{\mathrm{NA}}|/\langle R \rangle$ and (b) $D^2_{\mathrm{min}}/\langle R \rangle^2$, both as a function of the normalized particle size $R / \langle R \rangle$. Smaller particles have higher $|\Delta r_{\mathrm{NA}|}$ and $D^2_{\mathrm{min}}$ than larger particles.  This is because larger particles are more likely to follow the mean flow, which forces smaller particles to maneuver around these large particles.  To have enough data for a meaningful result, we average the observations of the three largest particles in C3 together.
}
\label{fig:D2min_persize}
\end{figure}

\begin{figure*}
\includegraphics[scale=0.59, origin=c]{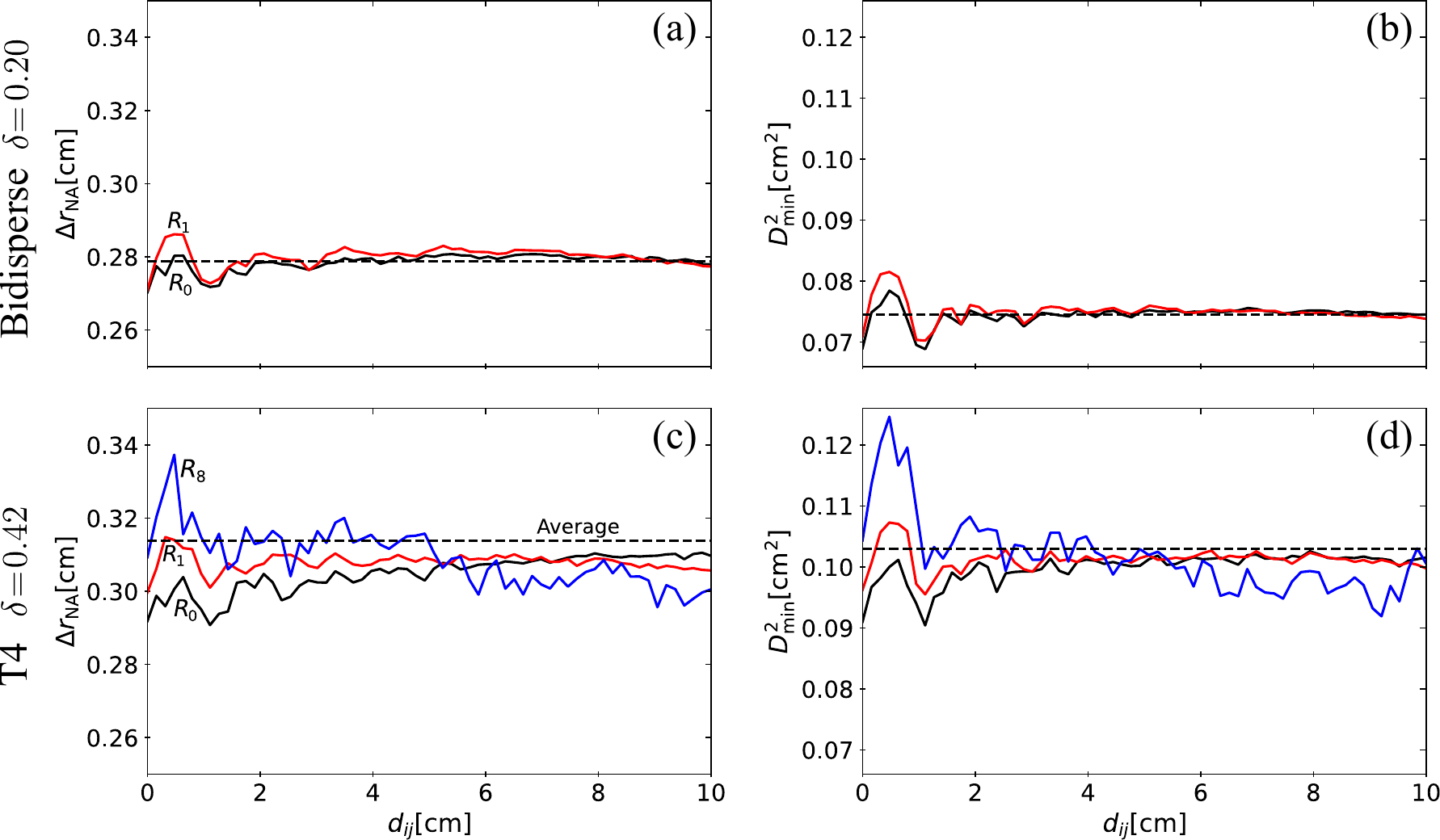}
\caption{\small $\Delta r_{\mathrm{NA}}$ is shown in (a,c) and $D^2_{\mathrm{min}}$ in (b,d), both as a function of distance($d_{0i}$) from the edge of a reference particle with radius $R$ to all other particles.  The top panels (a,b) correspond to the bidisperse distribution, and the bottom panels (c,d) to the T4 distribution. At $d_{0i}=0$, there's a valley for all cases, corresponding to particles in direct contact to the reference, resulting in low $\Delta r_{\mathrm{NA}}$ and $D^2_{\mathrm{min}}$, due to the tight packing. There's a another valley at $d_{0i}=2R_0$, which in this case corresponds to a buffer of one $R_0$ between the particles, but are otherwise in close packing, causing the minimum. In between these valleys we find a peak at $d_{ij}\sim 0.75 ~\mathrm{cm}$. These peaks are a good tool to measure the effect the reference particle has on its closest neighbors. This measurement is affected by particle size, an effect that can be seen for all distributions shown here, with larger particles having higher peaks. This can be seen by the differently colored curves. As shown in the legend, the blue curve correspond to the $R_8$ particles, while red and black to $R_0$ and $R_1$, respectively. The dashed line shows the average value for all particles.}
\label{fig:D2svsdist}
\end{figure*}

Figures \ref{fig:3x4}(e,f) show the particles shaded according to their value of $D^2_{\mathrm{min}}$, where darker colors indicate higher values. A similar behavior to Figs.~\ref{fig:3x4}(c,d) is observed for $D^2_\mathrm{min}$.  For the bidisperse case [Fig.~\ref{fig:3x4}(e)] the highest values for $D^2_{\mathrm{min}}$ occur close to the upper right corner where particles navigate the turn.  The region with the second highest levels of $D^2_{\mathrm{min}}$ is in the central zone.  In Fig.~\ref{fig:3x4}(f), corresponding to the T4 case, the larger particles are again seen to have an effect on the flow. 
Here, the particles with the largest $D^2_{\mathrm{min}}$ are small particles that are close to large particles, showing that large particles disrupt the displacements of their neighbors.  The proximity to the upper right corner appears less relevant.  The large particles, while enhancing the $D^2_{\mathrm{min}}$ of their neighbors, have a lower $D^2_{\mathrm{min}}$ value themselves.

Our goal is to understand the role of particle size, and Figs.~\ref{fig:3x4} (d,f) suggest that larger particles have less nonaffine motion. We know that the flow pattern is spatially heterogeneous, as will be discussed in more detail in Sec.~\ref{Itsmatrixtime}.  Nevertheless, we wish to find the average nonaffine motion as a function of particle size.  To do this, we pick a particle radius from a given experimental condition.  We calculate the values of $\Delta r_\mathrm{NA}/\langle R \rangle$ and $D^2_{\mathrm{min}}/\langle R \rangle^2$ for all particles of that size and all times.  We then find the mean values of these as a function of $(x,y)$, similar to how we find the flow field $\Delta \vec{r}_{\mathrm{mean}}(x,y)$.  Finally, we average the resulting fields over $(x,y)$.  We do this for all particle radii and all particle size distributions, with the results shown in Fig.~\ref{fig:D2min_persize}.  The one exception to this procedure is for the three largest particle sizes in the C3 particle size distribution, for which their small numbers do not give us adequate statistics.  Accordingly, we average the observations of these three particle sizes together to calculate the nonaffine motion as a function of $(x,y)$, and then plot the $(x,y)$ averaged results at the mean radius of the three particle sizes.  Figure~\ref{fig:D2min_persize} shows that the smaller particles have higher values of $\Delta r_{\mathrm{NA}}$ and $D^2_{\mathrm{min}}$, whereas the larger particles have smaller values. This points to the previously given explanation, where large particles move according to the average displacement field to which they are subjected but cause other particles either to have to detour around them or be bumped out of the way.

To study how large particles affect the motions of their neighbors, we measure the mean values of $\Delta r_{\mathrm{NA}}$ and $D^2_{\mathrm{min}}$ conditioned on the distance a particle has from a particle of a specific size.  These results are plotted as a function of the edge to edge distance $d_{ij}$ in Fig.~\ref{fig:D2svsdist}.  Figures \ref{fig:D2svsdist}(a,b) show the results for the bidisperse distribution. There is a minimum at contact between the particles, followed by a peak, and then a valley at $d_{ij}=2R_0$. These oscillations are more pronounced for the neighbors of particles with radius $R_1$, the larger of the two species in the bidisperse distribution [the top red curves in panels (a,b)].  Similar trends are stronger for the tridisperse distribution T4, shown in Fig.~\ref{fig:D2svsdist}(c,d).  Here, the largest particles (size $R_8$) strongly increase the nonaffine motion of their neighbors [the top blue curves in panels (c,d)]. These results confirm the conceptual picture sketched above that the large particles act as obstacles moving with the ``wrong'' velocity for some of their neighbors, forcing those neighbors to move nonaffinely.  For larger separations $d_{ij}$, the measures level out towards the average (albeit with noise).  The ``signal'' of the perturbation appears to be short-ranged and is within the noise for $d_{ij} \gtrsim 3$~cm, a distance equal to $2.4R_0 = 1.6R_1$ in terms of the two smallest particle sizes.  This short-ranged influence is comparable to that seen in a prior experiment which studied the oscillatory shear of emulsions \cite{clara-rahola_affine_2015}, although simulations of a 2D emulsion model found longer range influences out to approximately 5 particle diameters \cite{jiang_effects_2023}.

\begin{figure}
\includegraphics[scale=0.58,origin=c]{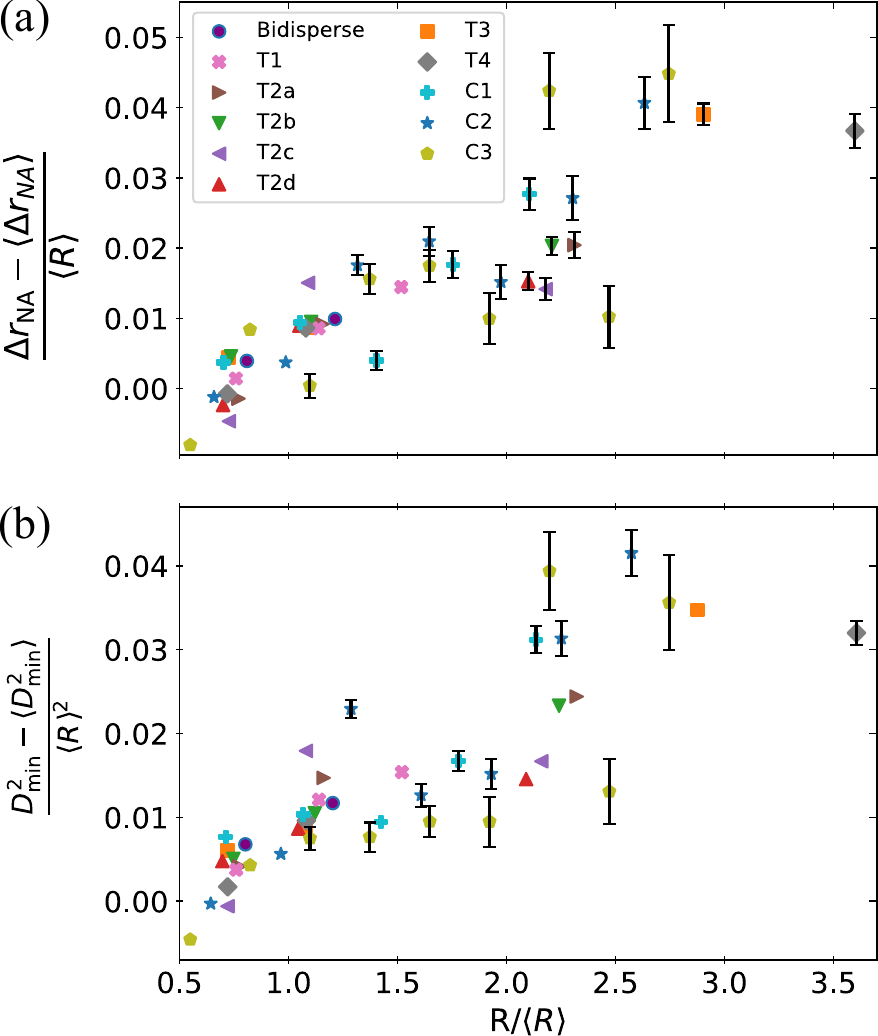}
\caption{\small These graphs show how particles of size $R/\langle R \rangle$ influence the motion of their neighbors.  (a) Peak ($\Delta r_{\mathrm{NA}}-\langle \Delta r_{\mathrm{NA}} \rangle)/\langle R \rangle$.  (b) Peak $(D^2_{\mathrm{min}}-\langle D^2_{\mathrm{min}} \rangle)/\langle R \rangle^2$.  The symbols correspond to distinct particle size distributions, given by the legend in (a).  The peak height is measured from data similar to that shown in Fig.~\ref{fig:D2svsdist}, where the peak is measured for $d_{0j}<2R_0$ from the reference particle.  Larger particles have stronger influences on their neighbors. Error bars are only shown when the error bar is larger than the scatter symbol.{\color{red}}}
\label{fig:NA_peaks_rads}
\end{figure}

The heights of the first peak in Figs.~\ref{fig:D2svsdist} are a good measure of the effect a particle has on the flow of its neighbors. We can then characterize this disturbance of flow caused by particles of size $R$ by calculating the difference in value between these peaks and the average $\Delta r_{\mathrm{NA}}$ and $D^2_{\mathrm{min}}$, for each particle size $R$, and for every size distribution.  Here we will nondimensionalize all lengths by the mean radius $\langle R \rangle$ for the relevant particle size distribution. Figure \ref{fig:NA_peaks_rads} shows the normalized differences in peaks for (a) $\Delta r_{\mathrm{NA}}$ and (b) $D^2_{\mathrm{min}}$ as a function of size $R$, and with the different symbols corresponding to the different size distributions studied. As hinted both by Figs.~\ref{fig:3x4} and Figs.~\ref{fig:D2svsdist}, the larger a particle, the larger the effect it will have on its neighbors, clearly shown in the growth of $\Delta r_{\mathrm{NA}}$ and $D^2_{\mathrm{min}}$ for larger $R/\langle R \rangle$.  The imperfect data collapse suggests there may be influences of the particle size distribution, although there is no clear trend.

The data shown in Figs.~\ref{fig:D2min_persize} and ~\ref{fig:NA_peaks_rads} show opposite trends as a function of $R$, and these opposite trends emphasize our conceptual story.  Larger particles are subjected to the mean flow of all of their surrounding neighbors, resulting in less nonaffine motion, confirmed in Fig.~\ref{fig:D2min_persize}.  These larger particles thus disrupt the flow of their neighboring smaller particles, forcing these smaller particles in a competition between following the mean flow and following the motion of their larger neighbor.  Thus, the larger particles cause more nonaffine motion for their neighbors, confirmed in Fig.~\ref{fig:NA_peaks_rads}.  These results agree with prior observations of softer particles \cite{clara-rahola_affine_2015,jiang_effects_2023}.

\begin{figure}
\includegraphics[scale=0.61, origin=c]{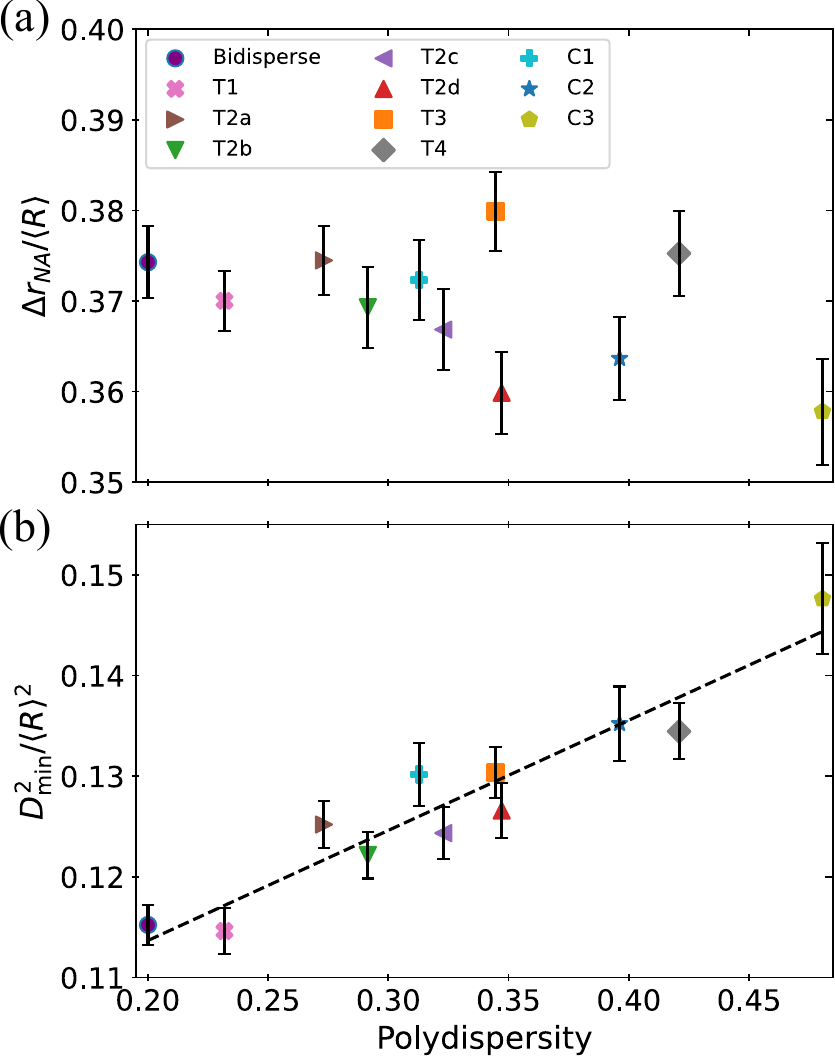}
\caption{\small (a) $\Delta r_{\mathrm{NA}}/\langle R \rangle$, and (b) $D^2_{\mathrm{min}}/\langle R \rangle ^2$, as a function of the polydispersity for each size distribution. For  $\Delta r_{\mathrm{NA}}/\langle R \rangle$, we see no relation to polydispersity, averaging to a value of $0.37$. On the other hand $D^2_\mathrm{min}/\langle R \rangle^2$ shows a positive relation with polydispersity.  The dashed line is a least-squares fit to the data; see text for details.}
\label{fig:D2min_vs_poly15}
\end{figure}

Our story focuses on the larger particles and it is plausible that the larger those particles are, the more strongly the overall particle motion is affected.  We test this conjecture by calculating the mean values of the spatial averaging of $\Delta r_{\mathrm{NA}}$ and $D^2_{\mathrm{min}}$ for all particles as a function of the polydispersity $\delta$ of the corresponding size distribution, plotted in Fig.~\ref{fig:D2min_vs_poly15}.  Surprisingly, $\Delta r_{\mathrm{NA}}/\langle R\rangle$ is not affected by changes in polydispersity.  The values for the distributions studied average to $\langle \Delta r_{\mathrm{NA}}/\langle R\rangle \rangle =0.37 \pm 0.01$.  Comparing this to the data shown in Fig.~\ref{fig:D2min_persize}, it appears that the smaller nonaffine motion for the few larger particles is balanced by the increased nonaffine motion of the more numerous smaller particles.  On the other hand, Fig.~\ref{fig:D2min_vs_poly15}(b) shows that $D^2_{\mathrm{min}}/\langle R \rangle ^2$ has a positive relation with polydispersity.  We fit the data using
\begin{equation}
D^2_{\mathrm{min}}/\langle R \rangle ^2=m\delta +b
\end{equation}
with slope $m=0.11$ and intercept $b=0.09$.  We note that this behavior is affected primarily by the polydispersity of a size distribution and not the size ratio itself, as evidenced by T2 family of distributions. All T2 distributions have the same size ratio but their $\langle D^2_\mathrm{min}\rangle/\langle R \rangle ^2$ values correlate mainly to their polydispersity. Conversely, two distributions with similar polydispersity but different size ratios $R_{\rm max}/R_{\rm min}$, T3 and T2d, have similar values of $\langle D^2_\mathrm{min}\rangle/\langle R \rangle ^2$.

\subsection{Time scale dependence}

\begin{figure}
\includegraphics[scale=0.62, origin=c]{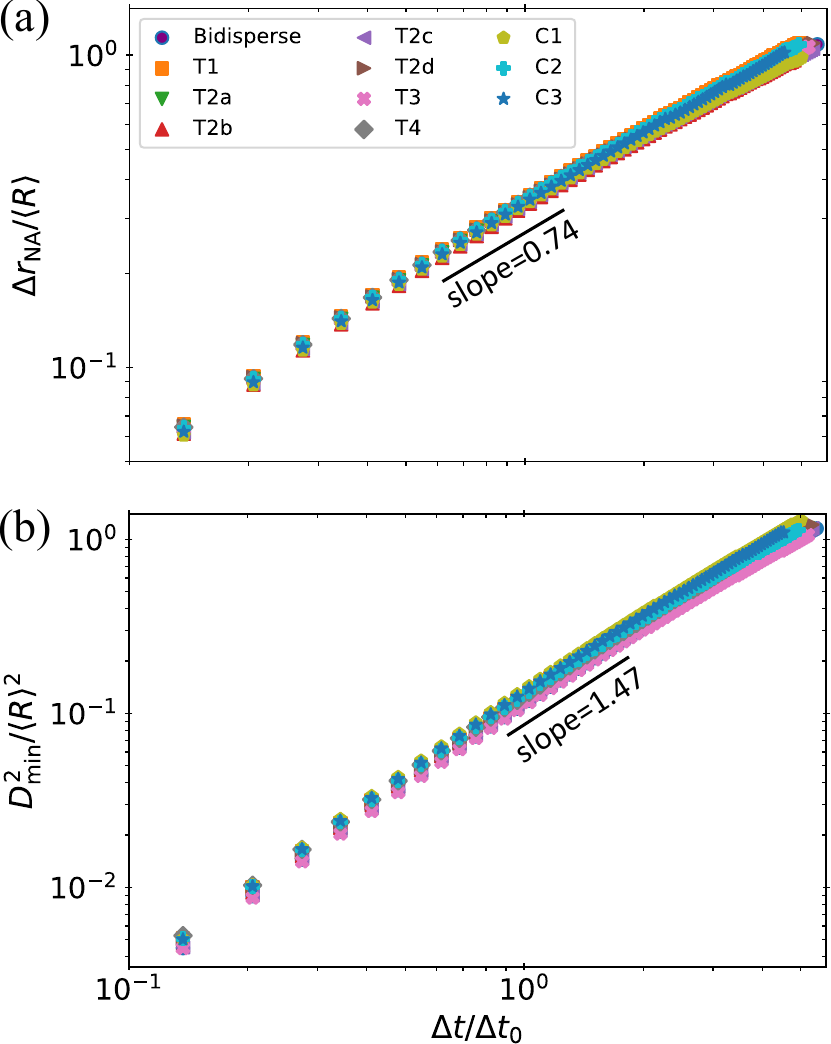}
\caption{\small Average $\Delta r_{\mathrm{NA}}/\langle R \rangle$, Fig.(a), and $D^2_{\mathrm{min}}/\langle R \rangle^2$, panel Fig.(b), for the central zone for various time intervals, on log-log scale, for all size distributions. The lines indicate power-law scaling with exponents as labeled.  All distributions share similar exponents and their corresponding curves do not cross each other for the observed $\Delta t$.}
\label{fig:DTloglog}
\end{figure}

All of the results above have used a set time scale $\Delta t_0 = 5$~s for calculating displacements.  To briefly investigate the influence of this choice, we study how the results of Fig.~\ref{fig:D2min_vs_poly15} depend on $\Delta t$.  We compute the average values for $\Delta r_{\mathrm{NA}}/\langle R \rangle$ and $D^2_{\mathrm{min}}/\langle R \rangle ^2$ for different $\Delta t/\Delta t_0$.  Here, to enhance the signal, we take the averages only over particles in the central region of the channel (the square region between the inner and outer corners of Fig.~\ref{fig:bkgrnd_flow}).  The data are plotted in Fig.~\ref{fig:DTloglog} for all time intervals.  All size distributions show a similar growth in values for increasing time intervals.  The data in Fig.~\ref{fig:DTloglog}(a) for $\Delta r_{\mathrm{NA}}/\langle R\rangle$ nearly superimpose on each other, which is expected as seen in Fig.~\ref{fig:D2min_vs_poly15}(a):  this measure of nonaffine motion is not sensitive to polydispersity, and this fact holds true for all $\Delta t$.  On the other hand, in Fig.~\ref{fig:DTloglog}(b) the $D^2_{\mathrm{min}}/\langle R \rangle ^2$ are slightly separated by polydispersity, in agreement with Fig.~\ref{fig:D2min_vs_poly15}(b).  The differences with polydispersity are more apparent in Fig.~\ref{fig:D2min_vs_poly15}(b); the logarithmic axis of Fig.~\ref{fig:DTloglog}(b) reduces the distance between the curves.

The time scale dependence of the nonaffine motion is well fit by power laws:
\begin{eqnarray}
    \Delta r_\mathrm{NA}/\langle R \rangle &\sim& \left( \frac{\Delta t}{\Delta t_0} \right) ^{\alpha_1}\\
    D^2_\mathrm{min}/\langle R \rangle ^2 &\sim& \left( \frac{\Delta t}{\Delta t_0} \right)^{\alpha_2}
\end{eqnarray}
with $\alpha_1 = 0.74$ and $\alpha_2 = 1.47$.  Given that $\Delta r_{\mathrm NA}$ has units of length and $D^2_{\mathrm min}$ units of length squared, it is reasonable that $\alpha_2 \approx 2 \alpha_1$.  Given that these power laws well-describe the data for each individual particle size distribution, we conclude that our observations of the character of the nonaffine motion are fairly robust over time scales $\Delta t$ up to the duration of our experiments.

\subsection{Strain clock}
\label{Itsmatrixtime}

For the bidisperse sample, we have noted throughout the previous subsection that there is more nonaffine motion near the top right corner, where there is more shearing.  Given that the strain rate is not spatially homogeneous, one can conjecture that the data are confounded by the specific locations of the particles.  To quantify this, we need to define a local strain rate.  
We can then compare the data which uses the fixed time interval $\Delta t_0=5$~s to define displacements and the data calculated using a fixed strain increment. We will first consider our fixed $\Delta t_0$ and examine the local strain rate.

\begin{figure*}
\includegraphics[scale=0.725, origin=c]{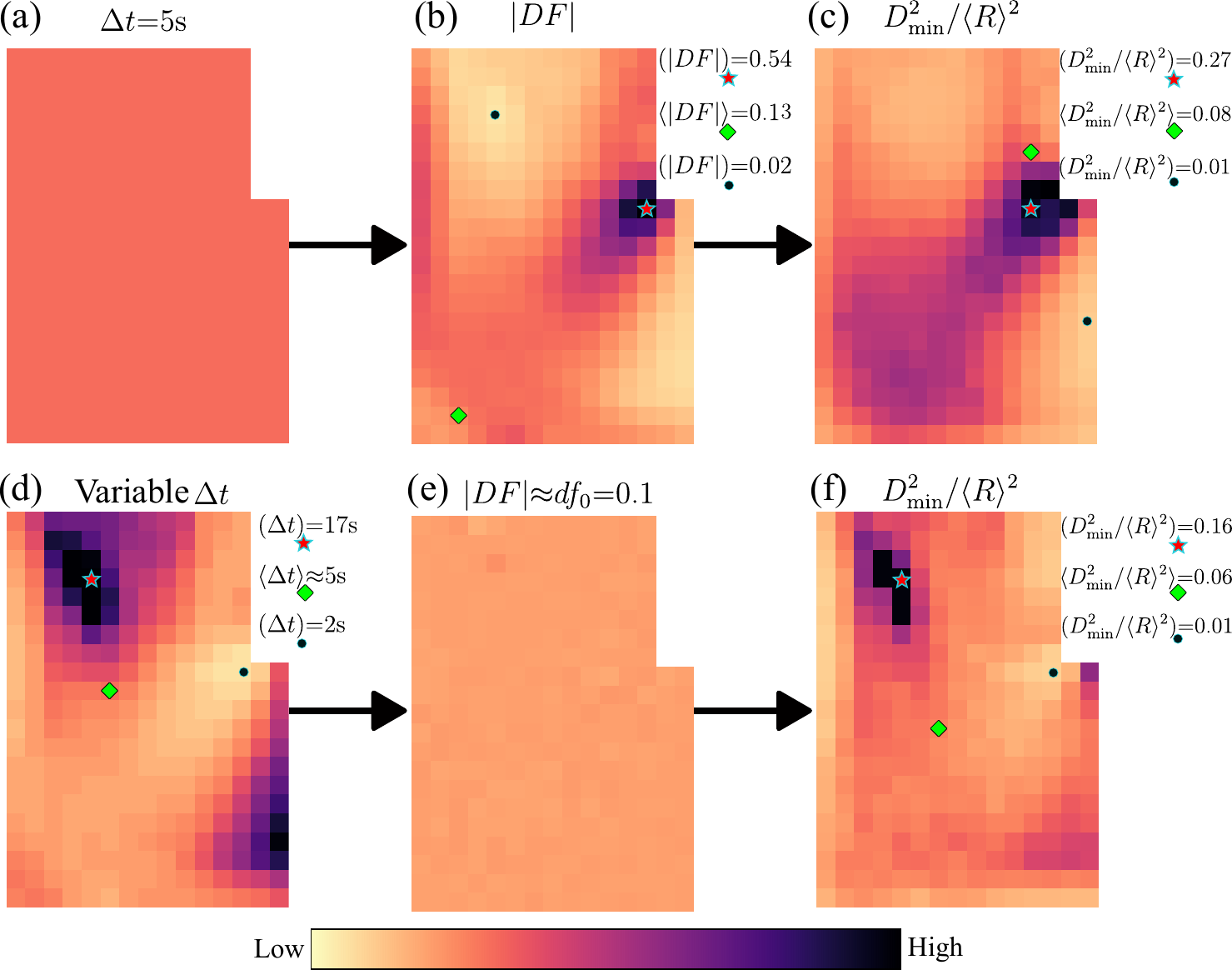}
\caption{\small Color maps for (a) $\Delta t= 5~\mathrm{s}$, (b) $|DF(\Delta t=5~\mathrm{s})|$, and (c) $D^2_{\mathrm{min}}(\Delta t=5~\mathrm{s})/\langle R \rangle ^2$ on the top row. On the bottom row we show the same quantities but with (d) $\Delta t(x.y)$, such that (e) $|DF(\Delta t(x,y))|\approx df_0$. (f) $D^2_\mathrm{min}(\Delta t(x,y))/\langle R \rangle ^2$.  We see in the top row that $D^2_{\mathrm{min}}$ is strongly correlated to $|DF(\Delta t=5~\mathrm{s})|$, while in the case for $\Delta t(x,y)$, considering fixed strain increments (e) lead to a somewhat more spatially homogeneous nonaffine motion pattern in the central area in (f).}
\label{fig:6matrices}
\end{figure*}

The starting point is the mean displacement field $\Delta \vec{r}_{\mathrm mean}(x,y,\Delta t)$ where now we explicitly include the $\Delta t$ dependence.  We then use the following equation to calculate the strain tensor:
\begin{equation}
    DF_{i,j}(x,y,\Delta t)=\frac{\partial \Delta r_i(\Delta t)}{\partial r_j}
    \label{eq:DF_vec}
\end{equation}
\noindent where $i,j$ are the spatial coordinate indices and $\Delta r_i(\Delta t)$ is the displacement in the $i$ coordinate for a time step $\Delta t$. This tensor contains information on the strain of the flow of the particles and is the non-symmetric version of the Cauchy strain tensor. To measure the total strain at a given position, we calculate the Frobenius norm of the strain tensor:
\begin{equation}
    |DF(x,y,\Delta t)|=\sqrt{\sum_{i,j}\left(\frac{\partial \Delta x_i(\Delta t)}{\partial x_j}\right)^2}.
    \label{eq:DF_mod}
\end{equation}
This then is a scalar which quantifies the amount of strain occurring over the time scale $\Delta t$ at each location $(x,y)$.

Figure \ref{fig:6matrices}(a-c) shows the results for fixed $\Delta t$.  Panel (a) shows $\Delta t$ and is uniform, reflecting that $\Delta t$ is constant.  Panel (b) shows that there are regions of high strain, especially near the upper right corner.  The lower right inlet region is not shown, as this region is mostly plug flow and uninteresting ($|DF| \approx 0$).  Panel (c) then shows that for fixed $\Delta t$, indeed much of the nonaffine motion measured by $D^2_{\mathrm min}/\langle R \rangle ^2$ is located in the regions with large $|DF|$.  The diagonal from lower left corner to upper right corner has more strain: as shown in Fig.~\ref{fig:bkgrnd_flow}, the velocity is changing direction and magnitude in this diagonal region.  Naturally, particles will be required to rearrange, and Fig.~\ref{fig:6matrices}(c) confirms that $D^2_{\rm min}/\langle R \rangle ^2$ is larger here.

To consider the case of constant strain interval rather than constant time interval, we return to the $|DF(x,y,\Delta t)|$ data.  We then define $\Delta t(x,y)$ through $|DF(x,y,\Delta t)| = df_0 = 0.1$, where $df_0$ is a small strain.  The choice of $df_0$ is somewhat arbitrary, but is chosen so that a good portion of $\Delta t(x,y)$ is a comparable order of magnitude to $\Delta t_0 = 5$~s.  $\Delta t(x,y)$ is quantized by our imaging rate (6 images per second), so in practice we find the $\Delta t(x,y)$ that minimizes the difference between $|DF(x,y,\Delta t)|$ and $df_0$.  $\Delta t(x,y)$ is shown in Fig.~\ref{fig:6matrices}(d), where a strong dependence on position is apparent.  Near the bottom right where there is plug flow, and near the top left where there is also a small region of plug-like flow, $\Delta t$ must be large to achieve any significant local strain.  By allowing $\Delta t$ to depend on the position, we achieve our goal $|DF| \approx df_0$, as shown in panel (e), with residual noise due to the quantization of $\Delta t$.

Next, we calculate $\Delta \vec{r}_{{\rm NA}}/\langle R \rangle$ and $D^2_{\rm min}/\langle R \rangle ^2$ for all particles and use the entire range of time scales $\Delta t$.  Finally, we examine $D^2_{\rm min}(x,y)/\langle R \rangle ^2$ where at each $(x,y)$ we use the data calculated with $\Delta t = \Delta t(x,y)$ to ensure the strain increment is $df_0$.  The results for $D^2_{\rm min}(x,y)/\langle R \rangle ^2$ are shown in Fig.~\ref{fig:6matrices}(f).  The resulting $D^2_{\rm min}$ is a smoother function of $(x,y)$.  


For much of the channel, this confirms that $D^2_{\rm min}/\langle R \rangle ^2$ is to an extent determined by the amount of strain that occurs at a given position.  The exception is the top left region, which has a patch where little strain occurs: the particles in this location tend to move in a group at constant velocity.  This causes $D^2_{\rm min}$ to be larger nearby and within this group.  To exclude this region from the subsequent analysis, we will restrict our attention to locations with $\Delta t \leq 17$~s.  This excludes the plug-flow region at the bottom right inlet location, as well as the center of the dark patch in Fig.~\ref{fig:6matrices}(d).

\begin{figure}
\includegraphics[scale=0.6, origin=c]{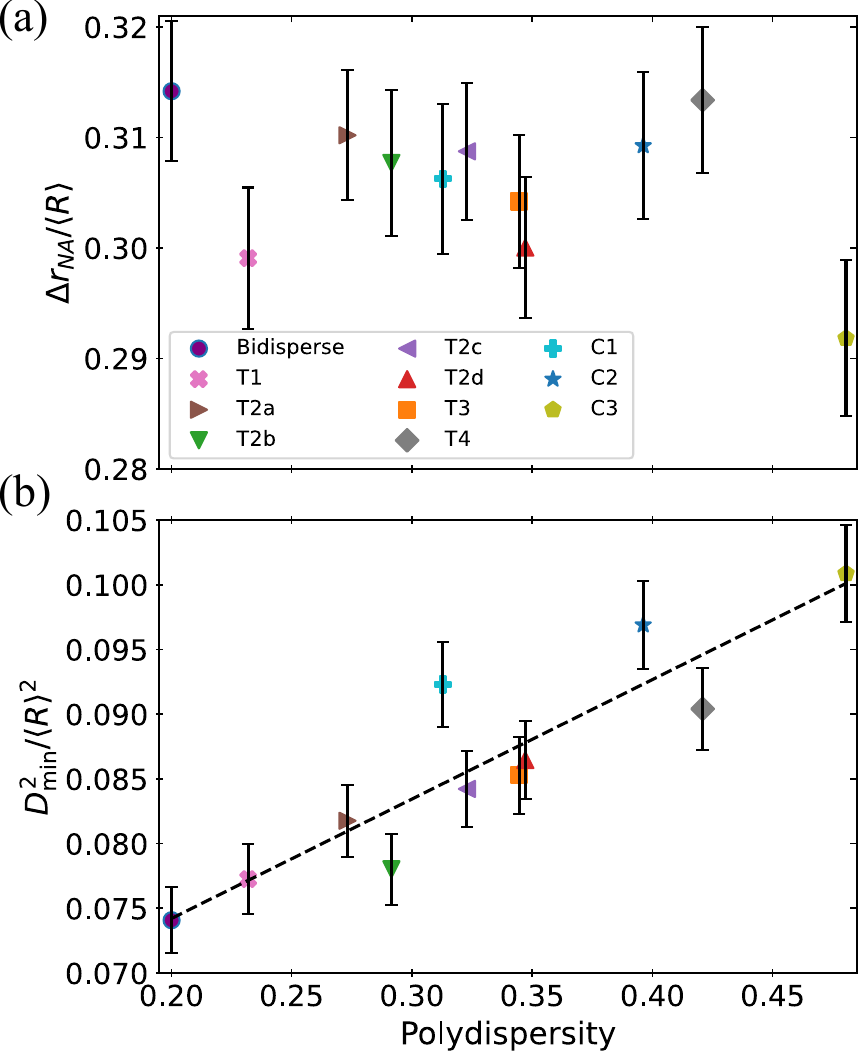}
\caption{\small  (a) $\Delta r_{\mathrm{NA}}/\langle R \rangle$ and (b) $D^2_{\mathrm{min}}/\langle R \rangle ^2$ as a function of polydispersity for all size distributions. In this figure $\Delta r_{\mathrm{NA}}/\langle R \rangle$ and $D^2_{\mathrm{min}}/\langle R \rangle^2$ are calculated for each particle using $\Delta t(x,y)$.  In contrast, Fig.~\ref{fig:D2min_vs_poly15} shows similar results using a fixed $\Delta t= 5~\mathrm{s}$. The dashed line in (b) is a least-squares fit to the data; see text for details.}
\label{fig:d2_v_poly_all}
\end{figure}

As just discussed, having $\Delta t(x,y)$ we can calculate $\Delta \vec{r}_{{\rm NA}}/\langle R \rangle$ and $D^2_{\rm min}/\langle R \rangle ^2$ for all particles based on a fixed strain increment $df_0$.  We then average over all particles within a given sample and plot these averages as a function of the polydispersity of the corresponding size distribution in Fig.~\ref{fig:d2_v_poly_all}. Similar to Fig.~\ref{fig:D2min_vs_poly15}(a), $\Delta r_{NA}/\langle R \rangle$ is not dependent on polydispersity.  Similar to Fig.~\ref{fig:D2min_vs_poly15}(b), $D^2_{\mathrm{min}}/\langle R \rangle^2$ shows a positive relation with polydispersity.  In this case we find $D^2_{\mathrm{min}}/\langle R \rangle^2 = m \delta + b$ with $m=0.09$.  The different value from the Fig.~\ref{fig:D2min_vs_poly15}(b) result are because the magnitude of $D^2_{\rm min}$ depends strongly on the choice of $\Delta t$ (or $df_0$), as shown in Fig.~\ref{fig:DTloglog}(b), so we do not expect a strict equivalence here.  The point, instead, is that analyzing the data using a fixed strain increment leads to a similar result as the analysis with a fixed time increment and shows that the spatial heterogeneity of our flow is not a critical confounding factor.


\section{Conclusion}

In this paper we have explored how polydispersity affects nonaffine displacement and particle rearrangement in granular flows.  
Consistent with previous work that studied similar systems \cite{clara-rahola_affine_2015,jiang_effects_2023}, we find that large particles tend to move similarly to mean flow, as they average over the forces from the many discrete particles they are contacting.  This then disrupts the flow of smaller neighboring particles, which need to navigate around the larger particles, thus causing the smaller particles to move nonaffinely.

These observations are true even for the bidisperse case, highlighting that even when the two particle sizes are quite similar (size ratio $1:1.5$ in our case), there is nonetheless a measurable difference in their nonaffine motion.  Increasing the polydispersity of the particle size distribution quantitatively increases the observable effects.  As polydispersity changes from 0.20 to 0.48, $D^2_{\mathrm min}/\langle R \rangle^2$ increases by nearly 30\% [Fig.~\ref{fig:d2_v_poly_all}(b)].  For broad size distributions with particle sizes varying by a factor of 5, Fig.~\ref{fig:D2min_persize} shows the largest particles have on average a magnitude of nonaffine motion $|\Delta \vec{r}_{mathrm NA}$ that is 16\% smaller than that of the smallest particles; and likewise while the data are noisier, $D^2_{\mathrm min}$ is smaller for the larger particles.  Finally, Fig.~\ref{fig:NA_peaks_rads} shows that the smallest particles barely perturb the motion of their neighbors, whereas the largest particles significantly enhance the nonaffine motion in their immediate vicinity.  The range of this enhancement is fairly short, about 2-3 small particle diameters.

Our analysis shows some differences between globally nonaffine motion ($\Delta \vec{r}_{\mathrm NA}$) and locally nonaffine motion ($D^2_{\rm min}$).  The former compares particle motion to the spatially smooth mean flow, whereas the latter compares particle motion to a flow defined locally in space and time.  The globally nonaffine motion is not significantly influenced by the particle polydispersity, suggesting that enhanced nonaffine motion for smaller particles is balanced by a decreased nonaffine motion for the larger particles.  The locally nonaffine motion has the dependence on sample polydispersity.  Both types of nonaffine motion are strongly enhanced in the presence of large particles.  As noted in prior work, this implies that mixing can be enhanced in these mixtures of particle sizes \cite{jiang_effects_2023}.

We also see that while large and small particles play different roles, it appears that polydispersity is the most significant factor determining the results; the size ratio $R_{\rm max}/R_{\rm min}$ between the largest and the smallest particles matters less. This is seen in the comparison of tridisperse distributions with fixed particle sizes but differing polydispersity, where polydispersity changes the results in a predictable way.  In contrast, data from size distributions with different $R_{\rm max}/R_{\rm min}$ but matched polydispersity have essentially equivalent results.  

In summary, we find that the flow of highly polydisperse materials is dramatically more complex than the flow of less polydisperse materials.  This suggests that models of localized rearrangements in the flow of amorphous materials may need to be adjusted to account for the roles of particle size and overall polydispersity \cite{falk_dynamics_1998,desmond15}.  Not all particles are equivalent; not all particle size distributions are equivalent.

\begin{acknowledgments}

This material is based upon work supported by the National Science Foundation under Grant Nos. CBET-1804186 and CBET-2306371. The authors also thank D.~J.~Meer and W.~Paliwal for helpful discussions.

\end{acknowledgments}


\bibliography{Citations}

\end{document}